\documentclass{SCIS2025}
\usepackage{amsmath}
\usepackage{multirow}
\usepackage{multicol}
\usepackage{makecell}
\usepackage{threeparttable}
\usepackage{lscape}
\usepackage{bm}

\newcommand{\circled}[1]{\textcircled{\raisebox{-.9pt}{#1}}}

\newcommand{\ours}{PromptCS}

\usepackage[normalem]{ulem}

\usepackage{pifont}
\usepackage{tcolorbox}
\newcommand{\summary}[1]{
\begin{center}
\begin{tcolorbox}[colback=gray!15, colframe=black, boxsep=-0.15cm, middle=-0.15cm]
\textbf{\ding{46} Summary}
$\blacktriangleright$
{#1}
$\blacktriangleleft$
\end{tcolorbox}
\end{center}
}

\begin{document}
\ArticleType{RESEARCH PAPER}
\Year{2025}
\Month{}
\Vol{}
\No{}
\DOI{}
\ArtNo{}
\ReceiveDate{}
\ReviseDate{}
\AcceptDate{}
\OnlineDate{}
\AuthorMark{}
\AuthorCitation{}

\title{A Prompt Learning Framework for Source Code Summarization}{A Prompt Learning Framework for Source Code Summarization}

\author[1]{Tingting Xu}{}
\author[2]{Yun Miao}{}
\author[2]{Chunrong Fang}{}
\author[2]{Hanwei Qian}{}
\author[3]{Xia Feng}{{xiafeng@cityu.edu.mo}}
\author[1]{Zhenpeng Chen}{}
\author[1]{\\Chong Wang}{}
\author[1]{Jian Zhang}{}
\author[1]{Weisong Sun}{}
\author[2]{Zhenyu Chen}{}
\author[1]{Yang Liu}{}


\address[1]{College of Computing and Data Science, Nanyang Technological University, Singapore {\rm 639798}, Singapore}
\address[2]{State Key Laboratory for Novel Software Technology, Nanjing University, Nanjing {\rm 210093}, China}
\address[3]{Faculty of Data Science, City University of Macau, Macau {\rm 999078}, China}

\abstract{(Source) code summarization is the task of automatically generating natural language summaries (also called comments) for given code snippets. Recently, with the successful application of large language models (LLMs) in numerous fields, software engineering researchers have also attempted to adapt LLMs to solve code summarization tasks. The main adaptation schemes include instruction prompting, task-oriented (full-parameter) fine-tuning, and parameter-efficient fine-tuning (PEFT). However, instruction prompting involves designing crafted prompts and requires users to have professional domain knowledge, while task-oriented fine-tuning requires high training costs, and effective, tailored PEFT methods for code summarization are still lacking.

In this paper, we propose an effective prompt learning framework for code summarization called \ours{}. It no longer requires users to rack their brains to design effective prompts. Instead, \ours{} trains a prompt agent that can generate continuous prompts to unleash the potential for LLMs in code summarization. Compared to the human-written discrete prompt, the continuous prompts are produced under the guidance of LLMs and are therefore easier to understand by LLMs. \ours{} is non-invasive to LLMs and freezes the parameters of LLMs when training the prompt agent, which can greatly reduce the requirements for training resources. 
We evaluate the effectiveness of \ours{} on the CodeSearchNet dataset involving multiple programming languages. Experimental results show that \ours{} significantly outperforms instruction prompting schemes (including zero-shot learning and few-shot learning) on all four widely used metrics, including BLEU, METEOR, ROUGE-L, and SentenceBERT, and is comparable to the task-oriented fine-tuning scheme. 
In some base LLMs, e.g., CodeGen-Multi-2B and StarCoderBase-1B and -3B, \ours{} even outperforms the task-oriented fine-tuning scheme. More importantly, the training efficiency of \ours{} is faster than the task-oriented fine-tuning scheme, with a more pronounced advantage on larger LLMs. The results of the human evaluation demonstrate that \ours{} can generate more good summaries compared to baselines.
}


\maketitle

\section{Introduction}
\label{sec:introduction}
Source code comments play a critical role in facilitating program comprehension~\cite{1988-Program-Readability, 2020-Human-Study-Code-Summarization, 2023-EACS} and software maintenance~\cite{2005-Documentation-Essential-Software-Maintenance, 2023-Automatic-Code-Summarization-via-ChatGPT}. However, existing research~\cite{2018-TL-CodeSum, 2022-Practitioners-Expectations-on-Comment-Generation} demonstrates that lack of high-quality code comments is a common problem in the software industry. In addition, comments are often absent, unmatched, and outdated during software evolution~\cite{2005-Documentation-Essential-Software-Maintenance}. These practical problems drive the research on source code summarization. Source code summarization (code summarization for short) is the task of automatically generating code summaries (i.e., comments). Over the past decade, it has always been one of the research hotspots in the field of software engineering~\cite{2010-Program-Comprehension-with-Code-Summarization, 2013-Evaluating-Source-Code-Summarization, 2018-DeepCom, 2020-Rencos, 2020-Human-Study-Code-Summarization, 2022-Evaluation-Neural-Code-Summarization, 2023-Function-Call-Graph-Context-Encoding-Code-Summarization, 2023-Adapter-Tuning-Code-Search-and-Summarization, 2023-Automatic-Code-Summarization-via-ChatGPT, 2023-Chatgpt-Programming-Assistant}.

Recently, with the success of large language models (LLMs) in natural language processing (NLP)~\cite{2022-Shortcut-Learning-of-LLM-in-NLU, 2023-ChatGPT-General-Purpose-NLP-Solver}, an increasing number of software engineering (SE) researchers have started integrating them into the resolution process of various SE tasks~\cite{2021-Evaluating-LLM-Trained-on-Code, 2023-LLM-for-SE, 2023-LLMs-for-SE-A-Literature-Review}. 
Similar to LLMs for NLP (e.g., ChatGPT~\cite{2022-ChatGPT} and LLaMA~\cite{2023-LLaMA}) tasks, there are many LLMs of code for SE tasks, e.g., Codex~\cite{2023-Codex}, StarCoder~\cite{2023-StarCoder}, CodeGen~\cite{2023-CodeGen}, and PolyCoder~\cite{2022-Systematic-Evaluation-of-Code-LLMs}. 
In this paper, we pay more attention to the application of LLMs on code summarization tasks. Existing studies demonstrate that LLMs trained on massive corpora of texts have shown their particularly exciting ability to perform new tasks from textual instructions (i.e., in zero-shot learning setting) or from a few examples (i.e., in few-shot learning setting)~\cite{2020-Language-Models-Few-Shot-Learners, 2022-Few-shot-Training-LLMs-for-Code-Summarization}. Therefore, there have been several recent studies investigating the effectiveness of instruction prompting with zero-shot and few-shot learning in adapting LLMs to code summarization tasks. For example, Sun et al.~\cite{2023-Automatic-Code-Summarization-via-ChatGPT} evaluate ChatGPT's performance on zero-shot code summarization tasks. They design several heuristic questions/instructions to collect the feedback of ChatGPT, thereby finding an appropriate prompt to guide ChatGPT to generate in-distribution code summaries. However, their experimental results on the large-scale CSN-Python dataset show that this prompt fails to induce ChatGPT to generate satisfactory summaries. We utilize their prompt to guide ChatGPT in generating summaries and the result is unsatisfactory compared to the result of task-oriented fine-tuning. It indicates that even professional researchers in the field of code summarization have difficulty designing good prompts. 
Ahmed et al.~\cite{2022-Few-shot-Training-LLMs-for-Code-Summarization} investigate the effectiveness of the instruction prompting with few-shot learning in adapting LLMs to code summarization tasks. To discover the appropriate example number for few-shot learning, they try multiple sets of samples with different numbers (including 5, 10, and 15) on a small-scale test dataset. Then, they utilize 10 samples to conduct instruction prompting on Codex and find it can outperform fine-tuned foundation models (e.g., CodeT5~\cite{2021-CodeT5}). We utilize the same 10 samples to conduct instruction prompting on open-source LLMs (e.g., StarCoderBase-3B) and find that their performance is still much worse than fine-tuned LLMs, detailed in Section~\ref{subsubsec:answer_to_RQ1}. 
This phenomenon indicates that for few-shot learning, which samples to choose and how many samples to choose require users to make decisions based on professional knowledge and continuous trial, as demonstrated by the work~\cite{2023-What-Makes-Good-In-Context-Demonstrations}.

Another straightforward scheme to adapt LLMs to better accomplish downstream tasks is task-oriented (full-parameter) fine-tuning~\cite{2023-InferFix, 2021-NASOA}. Task-oriented fine-tuning updates the weights of LLMs by training on thousands of supervised labels specific to the desired task. The main advantage of fine-tuning is strong performance on the task benchmark~\cite{2020-Language-Models-Few-Shot-Learners, 2019-Language-Models-Unsupervised-Multitask-Learners, 2022-Finetuned-Language-Models-are-Zero-Shot-Learners}. 
For example, Jin et al.~\cite{2023-InferFix} fine-tune Codex with 12 billion (B) on supervised bug-fix data. Although fine-tuning significantly enhances the model's ability in bug fixing, it comes with a high training cost. 
For instance, to conduct full model fine-tuning (updating all weights of Codex), Jin et al. build a training environment consisting of 64 32GB V100 GPUs, which is beyond the reach of many research teams. In this paper, we also experiment with the task-oriented fine-tuning scheme on code summarization tasks, which similarly incurs a high training cost, detailed in Section~\ref{subsubsec:answer_to_RQ1}. As the size of LLMs continues to grow, the cost of fine-tuning can be quite substantial. 
In recent years, researchers in NLP have proposed numerous parameter-efficient fine-tuning (PEFT) methods to reduce the cost of adapting LLMs, such as P-tuning~\cite{2022-P-Tuning}, Prefix-tuning~\cite{2021-Prefix-Tuning}, and LoRA~\cite{2022-LoRA}. PEFT methods can achieve comparable performance to task-oriented fine-tuning by updating fewer parameters~\cite{2023-PEFT-of-Pre-trained-LMs, 2022-Few-Shot-PEFT-is-Better-ICL, 2022-P-Tuning}. Inspired by these works, many SE researchers have also begun investigating the performance of PEFT methods for adapting LLMs to code-related tasks~\cite{2024-PEFT-in-Code-Change-Learning, 2023-Empirical-Study-of-PEFT-for-Code-Models, 2023-Parameter-Efficient-Fine-Tuning-Code-Generation, 2022-No-More-Fine-tuning-in-Code-Intelligence}, detailed in Section~\ref{subsec:llm_for_se}.
Most of these works are empirical studies; although their experiments cover various PEFT methods and a range of code-related tasks, they lack in-depth analysis of the performance of specific PEFT methods and the factors influencing their effectiveness on particular code-related tasks. Consequently, effective, customized PEFT methods for code summarization tasks are still lacking. 

In this paper, we propose a prompt learning framework specifically for code summarization, called \ours{}. Like P-tuning and Prefix-tuning, the key feature of \ours{} is its ability to free developers from the need to manually design intricate prompts. Specifically, \ours{} devises two collaborating components to achieve this feature, including a prompt agent and an LLM. The prompt agent is responsible for generating the continuous prompt that can induce the capacity of the LLM to perform code summarization tasks. 
The core of the prompt agent is a deep learning (DL) based prompt encoder, which takes a pseudo prompt consisting of $n$ learnable tokens as input and produces a prompt embedding (i.e., continuous prompt).  
The prompt agent is trained under the guidance of the LLM. Therefore, the well-trained prompt agent can produce a continuous prompt that is more suitable for the LLM than the human-written discrete prompt. More importantly, unlike the task-oriented fine-tuning scheme that changes the parameters of the LLMs, \ours{} is non-invasive to LLMs. 
\ours{} freezes the parameters of LLMs during the training process and only updates the parameters of the prompt agent, which can greatly reduce the requirements for training resources. Additionally, the non-intrusive approach offers another advantage: multiple downstream tasks can utilize the same base LLM, requiring only task-specific prompt agents to be trained. These prompt agents do not alter the parameters of the LLM, preserving its performance on other tasks. When new downstream tasks arise, a new prompt agent can simply be trained without impacting the LLM's effectiveness on existing tasks.

In summary, we make the following contributions.
\begin{itemize}

    \item We propose a novel prompt learning framework for code summarization called \ours{}. \ours{} is able to generate high-quality summaries for input code snippets. In addition, \ours{} is a general framework and can be combined with multiple LLMs (see experimental results in Section~\ref{subsubsec:answer_to_RQ3}). We open source the code of \ours{}~\cite{2023-Artifacts-of-PromptCS} to facilitate future research and application.

    \item We conduct extensive experiments on a widely used benchmark dataset to evaluate \ours{}. Experimental results show that in terms of all four metrics (i.e., BLEU, METEOR, ROUGE-L, and SentenceBERT), \ours{} significantly outperforms the instruction prompting schemes with zero-shot and few-shot learning, and is comparable to the task-oriented fine-tuning scheme. On some LLMs, e.g., StarCoderBase-1B and StarCoderBase-3B, \ours{} even outperforms the task-oriented fine-tuning scheme. In terms of training cost, \ours{} is significantly lower than the task-oriented fine-tuning scheme. For example, when using the StarCoderBase-7B model as the base LLM, \ours{} takes about 67 hours to train the prompt agent, while the task-oriented fine-tuning scheme takes about 211 hours for one epoch training, detailed in Section~\ref{subsubsec:answer_to_RQ1}. 

    \item We conduct a qualitative human evaluation to evaluate the summaries generated by \ours{} and baselines (including instruct prompting schemes with zero-shot and few-shot learning and the task-oriented fine-tuning scheme. The statistical results show that compared with baselines, the summaries generated by \ours{} are more similar to the ground-truth summaries (detailed in Section~\ref{subsubsec:human_evaluation}).
\end{itemize}

The rest of this paper is organized as follows. Section~\ref{sec:background} describes the background of code summarization and large language model.  
Section~\ref{sec:methodology} introduces the design of \ours{}. Section~\ref{sec:evaluation_and_analysis} presents the design of the experiments in detail and gives the details of experiment results and analysis. Section~\ref{sec:case_study} presents some code summarization cases. 
Section~\ref{sec:related_work} discusses the related work of this paper. We conclude the paper in Section~\ref{sec:conclusion}.


\section{Background}
\label{sec:background}

\subsection{Code Summarization}
\label{subsec:code_summarization}

\begin{figure}[htbp]
  \centering
  \includegraphics[width=0.8\linewidth]{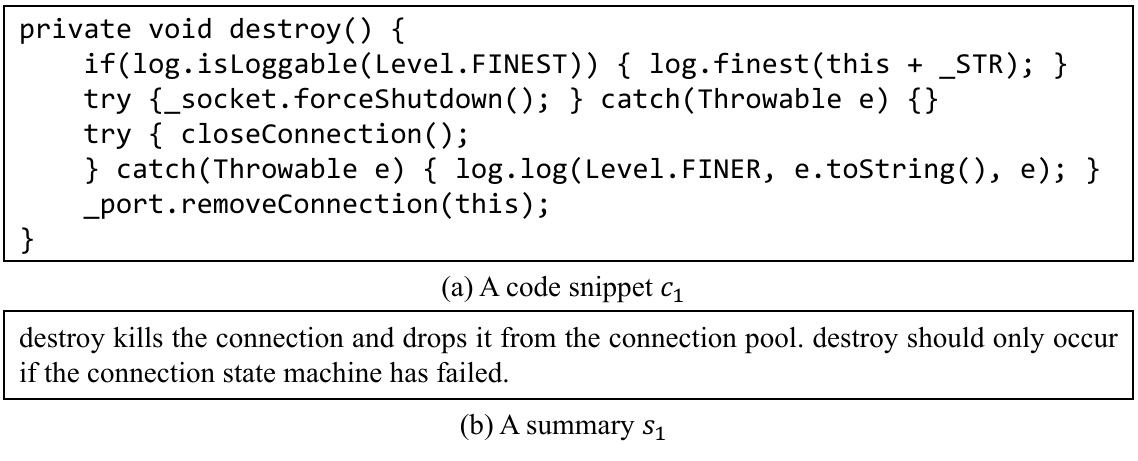}
  \caption{An example of a code snippet and its summary}
  \label{fig:example_of_query_and_code}
\end{figure}
 
Code summarization is the task of automatically generating natural language summaries for code snippets given by the developer. Such summaries are also called comments and explain the functionality of the code snippet~\cite{2019-Convolutional-Neural-Network-Code-Summarization, 2020-Hybrid-DeepCom, 2022-Practitioners-Expectations-on-Comment-Generation}. Figure~\ref{fig:example_of_query_and_code} shows an example. The code snippet $c_1$ in Figure~\ref{fig:example_of_query_and_code}(a) is provided by the developer. The summary ``destroy kills the connection and drops it from the connection pool. destroy should only occur if the connection state machine has failed.'' in Figure~\ref{fig:example_of_query_and_code}(b) is a summary that satisfies the developer's requirement. 
The research on code summarization can be traced back to as early as 2010 when Sonia Haiduc et al.~\cite{2010-Program-Comprehension-with-Code-Summarization} introduced automated text summarization technology to summarize source code to support program comprehension. 
Later on, following the significant success of neural machine translation (NMT) research in the field of NLP~\cite{2014-GRU, 2015-NMT-Jointly-Learning-to-Align-Translate}, a large number of researchers migrate its underlying DL-based encoder-decoder architecture to code summarization tasks~\cite{2020-Rencos, 2020-Code-to-Comment-Translation, 2020-Transformer-based-Approach-for-Code-Summarization, 2022-Evaluation-Neural-Code-Summarization, 2023-EACS, 2023-Adapter-Tuning-Code-Search-and-Summarization, 2023-Function-Call-Graph-Context-Encoding-Code-Summarization}. In the past ten years, code summarization has always been one of the hot research directions in the field of software engineering. In this paper, similar to~\cite{2022-Few-shot-Training-LLMs-for-Code-Summarization, 2023-Automatic-Code-Summarization-via-ChatGPT, 2023-Chatgpt-Programming-Assistant}, we focus on adapting LLMs for code summarization tasks. 
More details of related works are discussed in Section~\ref{subsec:code_summarization_related_work}.

\subsection{Large Language Model}
\label{subsec:large_language_model}
Existing research~\cite{2020-Scaling-Laws-Language-Models} demonstrates that scaling pre-trained language models (PLMs), e.g., increasing model size, can enhance model capacity for solving downstream tasks. To distinguish the early small-scale PLMs, more people are now willing to refer to the current generative PLMs with very large-scale parameters (often in the billions) as large language models (LLMs). When utilizing LLMs to accomplish downstream tasks, three components play a crucial role: a tokenizer, an input embedding layer, and module blocks. The tokenizer is responsible for converting the input text into the index representation that the model can understand, serving as a bridge between the input text and the model. The primary function of the input embedding layers is to transform the index representation into vector representations (also called embeddings). This process may involve techniques like word embeddings~\cite{2014-Glove} and positional encodings~\cite{2017-Transformer} to capture vocabulary and positional information. These techniques enhance the input embedding layer's ability to transform discrete data into meaningful continuous embeddings, facilitating subsequent processing. In the module blocks, the neural network structure is carefully designed to efficiently capture the complex relationships within embeddings. These blocks typically consist of multiple layers, with each layer responsible for processing different levels of features and abstract representations.

Currently, LLMs can be mainly divided into three categories: autoregressive language models, masked language models, and encoder-decoder language models~\cite{2022-Systematic-Evaluation-of-Code-LLMs}. The autoregressive language models generate sequences based on the outputs from previous time steps. They are powerful for modeling the probability of sequences. In the code summarization task, the model recursively predicts each word to generate a code summary. The currently popular LLMs all belong to this category, such as GPT series~\cite{2022-GPT-API}, Codex~\cite{2023-Codex}, StarCoder~\cite{2023-StarCoder}, CodeGen~\cite{2023-CodeGen},  PolyCoder~\cite{2022-Systematic-Evaluation-of-Code-LLMs}, and CodeLlama~\cite{2023-Code-Llama}. The masked language model can transform a given sequence into effective representations. They are trained using masked language modeling, a popular bidirectional objective function that aims to predict masked text pieces based on the surrounding context. Most of the early PLMs fall into this category, such as CodeBERT~\cite{2020-CodeBERT}, UniXCoder~\cite{2022-UniXcoder}, and GraphCodeBERT~\cite{2021-GraphCodeBERT}. The encoder-decoder language models consist of a structure with two main components, i.e., encoder and decoder. The encoder is responsible for transforming the input sequence into an intermediate representation, while the decoder utilizes this intermediate representation to generate the output sequence. CodeT5~\cite{2021-CodeT5} and PLBART~\cite{2021-PLBART} are examples of such models in code. 
In recent times, the impact of LLMs has been truly transformative, revolutionizing both academic research and various industrial applications. In this paper, we focus on the application of autoregressive LLMs in software engineering, particularly in code summarization.

\section{Methodology}
\label{sec:methodology}

\subsection{Overview}
\label{subsec:overview}
\begin{figure}[!t]
  \centering
  \includegraphics[width=0.9\linewidth]{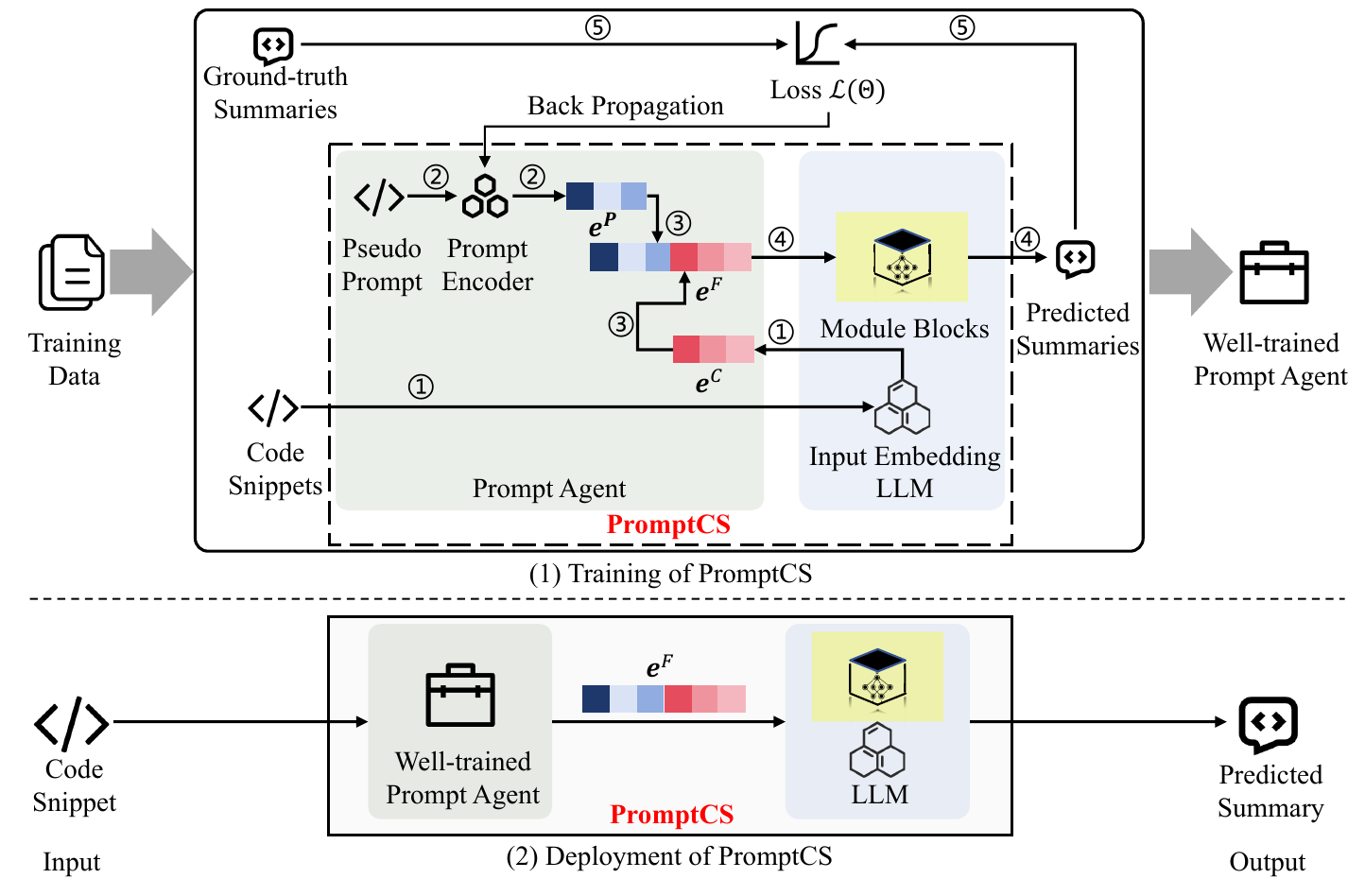}
  \caption{Overview of \ours{}}
  \label{fig:framework_of_our_approach}
\end{figure}

Figure~\ref{fig:framework_of_our_approach} illustrates the overview of \ours{}. The top part shows the training procedure of \ours{} and the bottom part shows the deployment/usage of \ours{} to support the code summarization service. \ours{} consists of two core components: a prompt agent and an LLM. The prompt agent is responsible for generating prompts that induce the LLM to generate natural language summaries for code snippets. 
\ours{} is non-invasive to LLMs, that is, the parameters of LLMs are always frozen and remain unchanged during the training procedure. Therefore, in the training procedure of \ours{}, our goal is to obtain a well-trained prompt agent. \ours{} utilizes the training data consisting of pairs of $\langle$code snippets, ground-truth summaries$\rangle$ to train the prompt agent. 
Specifically, \ours{} decomposes the training of the prompt agent into five steps. 
In step \circled{1}, \ours{} feeds code snippets to the LLM, utilizing its input embedding layer to generate the corresponding code embeddings denoted $\bm{e}^C$, detailed in Section~\ref{subsec:code_embedding_generation}. 
In step \circled{2}, \ours{} feeds a pseudo prompt into a prompt encoder to generate the prompt embedding denoted $\bm{e}^P$. The pseudo prompt is composed of $n$ learnable tokens with no actual meaning. 
The prompt encoder is a DL model and is trained along with the prompt agent. Details of this step are explained in Section~\ref{subsec:prompt_embedding_generation}. 
In step \circled{3}, \ours{} concatenates $\bm{e}^C$ and $\bm{e}^P$ together to produce fusion embeddings denoted $\bm{e}^{F}$. Similar to human-written discrete prompts, our \ours{} is non-invasive to input code snippets, that is, the prompt embedding does not break the integrity of the code embedding. Hence, $\bm{e}^P$ will only be concatenated to the front/back of $\bm{e}^C$, detailed in Section~\ref{subsec:fusion_embedding_generation}. $\bm{e}^{F}$ are then fed into the module blocks of the LLM to generate the predicted summaries (step \circled{4}). 
In step \circled{5}, based on the predicted summaries and the ground-truth summaries, \ours{} computes the loss and iteratively updates the model parameters of the prompt agent, detailed in Section~\ref{subsec:model_training}. 
Once the training phase is completed, the well-trained prompt agent is obtained. When \ours{} is deployed for usage, the prompt agent will automatically generate a prompt embedding for the code snippet given by the user to guide the LLM in generating a predicted summary. For the user, all of this happens seamlessly and is imperceptible.

\subsection{Code Embedding Generation}
\label{subsec:code_embedding_generation}
As mentioned in Section~\ref{subsec:large_language_model}, each LLM has three core components, i.e., a tokenizer, an input embedding layer, and module blocks. \ours{} directly employs the first two components of the LLM to accomplish the task of code embedding generation. 
Specifically, given a code snippet, our prompt agent first utilizes the tokenizer provided by the corresponding LLM to convert various elements in the code snippet, such as identifiers and symbols, into index representations. Each index representation corresponds to a token in the LLM vocabulary. Then, the prompt agent feeds the index representations to the input embedding layer of the LLM. The input embedding layer can encode the index representations to embeddings (i.e., $\bm{e}^C$) so that the module blocks can better understand their inner information.

\subsection{Prompt Embedding Generation}
\label{subsec:prompt_embedding_generation}
The function of the prompt embedding is to instruct the LLM to generate code summaries. 
As shown in Figure~\ref{fig:framework_of_our_approach}(1), the prompt embedding $\bm{e}^P$ is generated through the prompt encoder. 
The input of the prompt encoder is a pseudo prompt consisting of $n$ learnable tokens, denoted $p = \{t_{0}, t_{1}, \dots, t_{n}\}$, where $t_{i}$ refers to the $i$-th pseudo token. The role of pseudo tokens is to serve only as placeholders without carrying any actual meaning, and to indicate the length of the prompt used by the prompt encoder to generate the prompt embedding. 
The prompt encoder is a DL model capable of mapping $p$ to a continuous sequence of numbers, which is then input into the embedding layer to generate $\bm{T}=\{\bm{t}_{0}, \bm{t}_{1}, \dots, \bm{t}_{i}, \dots, \bm{t}_{n}\}$, where $\bm{t}_{i}$ is a trainable embedding tensor. This allows us to discover more effective continuous prompts that go beyond the original vocabulary of the LLM. 
Then, the prompt encoder applies a Bidirectional Long-short Term Memory network (BiLSTM), with a ReLU activated two-layer multilayer perceptron (MLP) to produce the real prompt embeddings $\bm{e}^P=\{\bm{e}_{0}, \bm{e}_{1}, \dots, \bm{e}_{n}\}$. The value $\bm{e}_{i}$ at the $i$-th position of $\bm{e}^P$ is calculated as follows:
\begin{equation}
    \footnotesize
    \bm{h}_{i}^f = LSTM(\bm{h}_{i-1}^f), \; \bm{h}_{i}^b = LSTM(\bm{h}_{i+1}^b), \; \bm{e}_{i} = MLP([\bm{h}_{i}^f, \bm{h}_{i}^b])
\end{equation}
where $\bm{h}_{i}^f$ represents the hidden state of the forward LSTM, while $\bm{h}_{i}^b$ represents the hidden state of the backward LSTM.

\subsection{Fusion Embedding Generation}
\label{subsec:fusion_embedding_generation}
\begin{figure}[htbp]
  \centering
  \includegraphics[width=\linewidth]{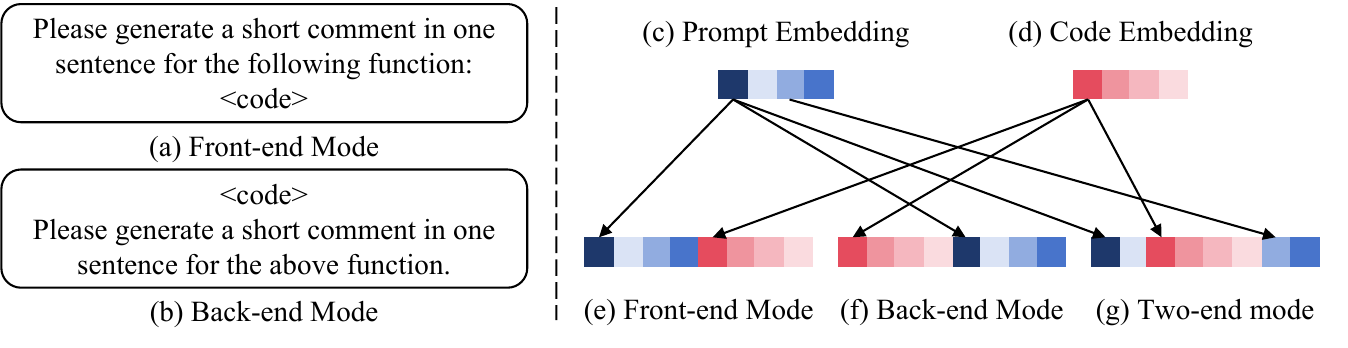}
  \caption{Example of prompt and code concatenation}
  \label{fig:example_of_concatenation}
\end{figure}

Similar to how humans concatenate the discrete prompt with the code snippet, \ours{} can also concatenate the continuous prompt generated by the prompt agent (i.e., prompt embedding) with the embedding of the code snippet in various ways. 
In order not to destroy the integrity of the code itself, a common human practice is to concatenate discrete prompts in front or back of the code snippet~\cite{2023-Chatgpt-Programming-Assistant, 2023-Automatic-Code-Summarization-via-ChatGPT}. 
The left side of Figure~\ref{fig:example_of_concatenation} shows an example of concatenating a discrete prompt and a code snippet (denoted as <code>), where (a) and (b) represent two modes respectively: front-end mode, concatenating the discrete prompt in front of the code snippet; and back-end mode, concatenating the discrete prompt in front of the code snippet.   
In this paper, in addition to following the aforementioned two modes, we also try a new mode called the two-end mode. In the two-end mode, the prompt embedding is split into two parts that will be concatenated in the front and back of the code embedding, respectively. 
This split is easy to implement. For example, we can split $\bm{e}^P= \{\bm{e}_0, \bm{e}_1, \dots, \bm{e}_n\}$ into $\bm{e}^P_1=\{\bm{e}_0, \bm{e}_1, \dots, \bm{e}_i\}$ and $\bm{e}^P_2=\{\bm{e}_{i+1}, \dots, \bm{e}_n\}$.
The right side of Figure~\ref{fig:example_of_concatenation} shows an example of concatenating a prompt embedding and a code embedding where (e), (f), and (g) showcase three concatenation modes, respectively.

\subsection{Model Training}
\label{subsec:model_training}
\noindent\textbf{Predicted Summary Generation.} 
In \ours{}, the predicted summary is generated by the second component, i.e., the LLM. As shown in Figure~\ref{fig:framework_of_our_approach}(1), \ours{} directly feeds the fusion embedding (i.e., $\bm{e}^F$) to the module blocks of the LLM that can generate natural language summaries. 
As mentioned in Section~\ref{subsec:large_language_model}, in this paper, we pay more attention to adapting autoregressive LLMs to the code summarization task. For autoregressive LLMs, predicting code summary can be viewed as a conditional generation task where the input is a context and the output is a sequence of tokens. 
Formally, let $\bm{z}=[\bm{e}^F, \bm{e}^{S}]$, where $\bm{e}^{S}$ is the summary embedding obtained by feeding the already generated summary into the input embedding layer of the LLM. And $\bm{z}_{i}$ denotes the $i$-th value in $\bm{z}$. The activation vector generated by the LLM at time step $i$ is $a_{i} = [a_{i}^{(1)}, \dots, a_{i}^{(n)}]$, which is a concatenation of the outputs of all activation layers at this time step. And $a_{i}^{(j)}$ is the activation vector of the $j$-th layer at time step $i$. The autoregressive LLM computes $a_{i}$ based on $z_{i}$ and the past activations $a_{<i}$ in its left context, as follows: $a_{i} = LLM(\bm{z}_{i}, a_{<i})$. 
Then, it utilizes a matrix $W$ to map the last layer of $a_{i}$ to logits $l_{i}$, which is the probability vector associated with the next word, i.e., $l_{i}= W \cdot a_{i}^{(n)}$. 
Finally, based on $l_{i}$, it further uses the function $P(\cdot)$ to select the token with the highest probability in the LLM vocabulary as the next token $s_{i+1}$ of the already generated summary, i.e., $s_{i+1} = P(l_{i})$.

\noindent\textbf{Train Process.}
During the training process of \ours{}, the parameters of LLM are frozen, and only the parameters of the prompt encoder are updated. Let $\hat{y}$ be the probability vector corresponding to the predicted summary and $y$ be the ground-truth summary. The loss function can be modeled as the following categorical cross-entropy loss function:
\begin{equation}
    \footnotesize
    \begin{aligned}
        & \mathcal{L}(\Theta) = -\sum_{i=1}^{C}y_{i}log\frac{exp(\hat{y_{i}})}{\sum_{j=1}^{C}exp(\hat{y_{j}})}
    \end{aligned}
\end{equation}
where $\Theta$ representing the trainable parameters of the model; $C$ is the number of tokens in the vocabulary. $\hat{y_{i}}$ and $y_{i}$ represent the probability of the predicted token and ground-truth token for each $i \in C$, respectively.
\section{Evaluation and Analysis}
\label{sec:evaluation_and_analysis}
We conduct a series of experiments to answer the following research questions (\textbf{RQs}):
\begin{itemize}
    \item \textbf{RQ1.} How effective is \ours{} in adapting LLMs to code summarization?

     \item \textbf{RQ2.} How do the key configurations, i.e., the prompt length and the concatenation mode of code and prompt embeddings, affect \ours{}?

    \item \textbf{RQ3.} How does the choice of the network architecture in the prompt encoder affect \ours{}?

    \item \textbf{RQ4.} How does training data size affect \ours{}?

    \item \textbf{RQ5.} How does \ours{} perform on code summarization tasks in other programming languages?

    \item \textbf{RQ6:} How does \ours{} perform in human evaluation?
\end{itemize}

\subsection{Experimental Setup}
\label{sec:experimental_setup}

\subsubsection{Dataset}
\label{subsubsec:dataset} We evaluate \ours{} on the CodeSearchNet (CSN) corpus~\cite{2019-CodeSearchNet-Challenge}, which is an extensive collection of code snippets accompanied by their comments written in six programming languages (including Go, Java, JavaScript, PHP, Python, and Ruby). This corpus is most commonly used in studying code summarization~\cite{2022-UniXcoder, 2022-Automatic-Source-Code-Summarization-With-GNN, 2022-Few-shot-Training-LLMs-for-Code-Summarization, 2023-EACS, 2023-Automatic-Code-Summarization-via-ChatGPT}. Considering that the original CSN corpus contains some low-quality data~\cite{2021-CodeXGLUE}, we use the dataset from the CodeXGLUE~\cite{2021-CodeXGLUE} code-to-text docstring generation task, which is built upon the CSN corpus and excludes defective data samples.

\subsubsection{Evaluation Metrics}
\label{subsubsec:evaluation_metrics} 
We leverage four metrics in the evaluation, including BLEU~\cite{2002-BLEU}, METEOR~\cite{2005-METEOR}, ROUGE~\cite{2004-ROUGE}, and SentenceBERT~\cite{2019-SentenceBERT}. These metrics are widely used in code summarization~\cite{2016-CODE-NN, 2018-Improving-Code-Summarization-via-DRL, 2018-TL-CodeSum, 2020-Rencos, 2022-Reinforcement-Learning-Guided-Code-Summarization, 2021-SiT, 2023-EACS}. 

\noindent\textbf{BLEU}, the abbreviation for BiLingual Evaluation Understudy~\cite{2002-BLEU}, is a variant of precision metric, which calculates the similarity by computing the n-gram precision of a generated summary to the ground-truth summary, with a penalty for the overly short length. In this paper, we follow~\cite{2021-SiT, 2023-EACS} and show the standard BLEU score which provides a cumulative score of 1-, 2-, 3-, and 4-grams.

\noindent\textbf{METEOR}~\cite{2005-METEOR} is introduced to address the concerns of using BLEU. METEOR combines n-gram precision and n-gram recall by taking their harmonic mean to compute a measure of similarity.

\noindent\textbf{ROUGE-L}, is a variant of ROUGE (Recall-oriented Understudy for Gisting Evaluation)~\cite{2004-ROUGE}. ROUGE-L is computed based on the longest common subsequence (LCS).

\noindent\textbf{SentenceBERT}~\cite{2022-Semantic-Metrics-for-Evaluating-Code-Summarization}. Unlike the above three metrics that mainly calculate the textual similarity between the ground truth summaries and the generated summaries, SentenceBERT measures the semantic similarity. SentenceBERT first converts the two compared summaries into embeddings in a unified vector space, and then uses cosine similarity to represent the semantic similarity between them. 

The scores of BLEU, ROUGE-L, METEOR, and SentenceBERT are in the range of [0, 1] and are reported in percentages. The higher the scores, the closer the generated summary is to the ground-truth summary, and the better the code summarization performance.

\subsubsection{Base LLMs}
\label{subsubsec:Base_LLMs}
We conduct experiments on four popular autoregressive LLMs, including three open-source LLMs: PolyCoder~\cite{2022-Systematic-Evaluation-of-Code-LLMs}, CodeGen-Multi~\cite{2023-CodeGen}, StarCoderBase~\cite{2023-StarCoder}, and a commercial LLM: ChatGPT~\cite{2022-ChatGPT}. 

\noindent\textbf{PolyCoder}~\cite{2022-Systematic-Evaluation-of-Code-LLMs} is released by researchers from Carnegie Mellon University. It is trained using the GPT NeoX toolkit on 249GB of code from 12 programming languages, and is available in three sizes: 160M, 0.4B, and 2.7B. 

\noindent\textbf{CodeGen-Multi}~\cite{2023-CodeGen} is released by Salesforce, which is a member of an autoregressive language models family for program synthesis. We use three sizes of CodeGen-Multi: 350M, 2B, and 6B.

\noindent\textbf{StarCoderBase}~\cite{2023-StarCoder} is jointly released by Hugging Face and ServiceNow in 2023. It is trained on 80+ programming languages from The Stack~\cite{2022-Stack}. 
And the renowned StarCoder is a fine-tuned version of StarCoderBase, trained on an additional 35 billion Python tokens. Considering that StarCoder is only available in a 16B version, which may not be suitable for conducting comprehensive comparison experiments. In our experiments, we use three sizes of StarCoderBase: 1B, 3B, and 7B.

\noindent\textbf{ChatGPT}~\cite{2022-ChatGPT} is possibly the most powerful LLM. To conduct experiments with its assistance, we utilize the OpenAI API, which is powered by a diverse set of models with different capabilities, such as GPT-3.5, DALL·E and Whisper. In our experiments, we use the gpt-3.5-turbo model, which distinguishes itself as the most capable and cost-effective option within the GPT-3.5 family.

\subsubsection{Experimental Settings}
\label{subsubsec:experimental_settings}
At the training stage, we set the mini-batch size to 16, and the learning rate to 5e-5. We employ the AdamW optimizer~\cite{2019-AdamW} along with a linear learning rate scheduler, following the default setup recommended by Hugging Face. To ensure comprehensive training, we employ an early stopping strategy based on the best validation BLEU to control the number of training epochs for the model. The early stopping patience is set to 4. The input sequences are padded to the maximum length with special tokens, which are selected based on the vocabulary of the LLM. 
To fine-tune StarCoderBase-7B and CodeGen-Multi-6B on a single A800, we utilize DeepSpeed 0.12.2. We set the Zero Redundancy Optimizer (ZeRO) to ZeRO-3 and enable the offloading of optimizer computation to CPU.
All models are implemented using the PyTorch 2.1.0 framework with Python 3.8. 
All experiments are conducted on a server equipped with one Nvidia A800 GPU with 80 GB memory, running on Ubuntu 20.04.4.

\subsection{Experimental Results}
\label{subsec:experimental_results}
In this section, we present and analyze the experimental results to answer the research questions.

\subsubsection{\textbf{RQ1: Effectiveness of \ours{}}}
\label{subsubsec:answer_to_RQ1}
\

\noindent1)\textit{\;Baselines:} To answer this research question, we compare our \ours{} to the following three schemes of adapting LLMs to code summarization tasks.

\noindent\textbf{Instruct Prompting (zero-shot).} This scheme directly uses human-written instructions to prompt LLMs to generate summaries for given code snippets. For StarCoderBase, CodeGen-Multi, and PolyCoder, we utilize the human-written prompt provided in the work~\cite{2023-LLM-for-Code-Explanation}. For ChatGPT, we utilize the human-written prompt provided in the work~\cite{2023-Automatic-Code-Summarization-via-ChatGPT}. Table~\ref{tab:RQ1-manual_prompt} presents the two human-written prompts mentioned above. The prompts we use have been proven effective in previous works. Considering that different prompts would threaten the fairness of RQ1, we also tried both prompt instructions (PI1 and PI2) for each LLM.

\begin{table}[!t]
    \caption{Manual prompts used in zero-shot learning to guide LLMs for code summarization}
    \centering
    \scriptsize
    \label{tab:RQ1-manual_prompt} 
    \begin{tabular}{ccl}
        \toprule
        
        Number & LLM & \multicolumn{1}{c}{Prompt Instruction} \\ 
        
        \midrule
        
        \multirow{3}{*}{PI1} & StarCoderBase & \multirow{3}{*}{\makecell[l]{//Human: You are a helpful code summarizer. Please describe in simple english the purpose of  \\ the following Java code snippet: $\langle$code$\rangle$ \\
        //Assistant:}} \\ 
        
        & CodeGen-Multi & \\ 
        
        & PolyCoder & \\ 
        
        \midrule
        
        PI2& ChatGPT & \makecell[l]{Please generate a short comment in one sentence for the following function: $\langle$code$\rangle$} \\ 
        
        \bottomrule
    \end{tabular}
\end{table}

\noindent\textbf{Instruct Prompting (few-shot).} In this scheme, we provide a few examples that demonstrate the nature of the task to let LLMs perform few-shot learning. 
We follow~\cite{2022-Few-shot-Training-LLMs-for-Code-Summarization, 2023-LLM-for-Code-Explanation} and provide 10 examples in the few-shot setting. Each example is a pair of $\langle code\;snippet, summary\rangle$ randomly selected from the training set.  
In practice, we directly leverage the 10 examples provided by Ahmed et al. in their GitHub repository~\cite{2022-Artifacts-of-Few-shot-Training-Code-Summarization}, since we use the same experimental dataset (i.e., the CSN corpus).

\noindent\textbf{Task-oriented Fine-tuning.} In this scheme, we perform the standard fine-tuning process and utilize all $\langle code\;snippets, summaries\rangle$ in the training set to fine-tune LLMs. During the fine-tuning process, all model parameters of LLMs may undergo changes.

\noindent\textbf{\ours{}.} This is the scheme proposed in this paper. \ours{} freezes the parameters of LLMs and only trains the prompt agent (i.e., prompt encoder). Similar to the task-oriented Fine-tuning scheme, we utilize all $\langle code\;snippets, summaries\rangle$ in the training set to train the prompt agent. Two key configurations of \ours{}, i.e., the prompt length and the concatenation mode of the code embedding and prompt embedding are set to 100 and back-end mode, respectively. More experiments on these two key configurations are discussed in Section~\ref{subsubsec:answer_to_RQ2}.

\begin{table*}[t]
    \caption{Effectiveness of \ours{} on the CSN-Java dataset. Zero-shot: Instruction Prompting with zero-shot learning; Few-shot: Instruction Prompting with few-shot learning; $\mathcal{B}$: BLEU; $\mathcal{M}$: METEOR; $\mathcal{R}$: ROUGE-L; $\mathcal{S}$: SentenceBERT.}
    \centering
    \scriptsize
    \tabcolsep=2pt
    \label{tab:effectiveness_of_PromptCS} 
    \begin{threeparttable}
        \begin{tabular}{cccccccccccccccccc}
        
            \toprule
            \multirow{2}{*}{LLM} & \multirow{2}{*}{\makecell{Model \\ Size}} & \multicolumn{4}{c}{Zero-shot} & \multicolumn{4}{c}{Few-shot} & \multicolumn{4}{c}{Task-oriented Fine-tuning} & \multicolumn{4}{c}{\ours} \\ 
            
             \cmidrule(lr){3-6} \cmidrule(lr){7-10} \cmidrule(lr){11-14}
             \cmidrule(lr){15-18}
            
             & & $\mathcal{B}$ & $\mathcal{M}$ & $\mathcal{R}$ & $\mathcal{S}$ & $\mathcal{B}$ & $\mathcal{M}$ & $\mathcal{R}$ & $\mathcal{S}$ & $\mathcal{B}$ & $\mathcal{M}$ & $\mathcal{R}$ & $\mathcal{S}$ & $\mathcal{B}$ & $\mathcal{M}$ & $\mathcal{R}$ & $\mathcal{S}$ \\


             \midrule
             
             \multirow{3}{*}{PolyCoder} & 160M & 7.98 & 3.40 & 13.82 & 16.12 & 13.76 & 8.85 & 27.58 & 49.11 & \textbf{17.79} & \textbf{12.92} & \textbf{36.24} & \textbf{59.21} & 16.01 & 11.68 & 34.04 & 56.44 \\
            
             & 0.4B & 7.18 & 2.67 & 11.55 & 16.89 & 14.71 & 10.16 & 29.89 & 5.21 & \textbf{18.80} & \textbf{13.29} & \textbf{37.14} & \textbf{59.88} & 16.62 & 12.18 & 34.84 & 57.61 \\
             
             & 2.7B & 7.68 & 2.28 & 12.23 & 12.92 & 13.80 & 8.82 & 27.48 & 49.26 & \textbf{19.33} & \textbf{13.66} & \textbf{37.97} & \textbf{60.70} & 18.73 & 13.00 & 37.24 & 59.87 \\ 
             
            \midrule
             
            \multirow{3}{*}{CodeGen-Multi} & 350M & 6.13 & 7.46 & 13.76 & 40.26 & 14.44 & 10.50 & 30.04 & 52.98 & \textbf{19.12} & \textbf{13.69} & \textbf{37.95} & \textbf{59.85} & 17.38 & 12.97 & 36.37 & 58.84 \\
            
            & 2B & 7.96 & 3.08 & 13.42 & 15.26 & 14.96 & 11.88 & 31.09 & 55.30 & 19.42 & 14.16 & 38.54 & 60.64 & \textbf{20.04} & \textbf{14.39} & \textbf{39.36} & \textbf{61.51} \\
            
            & 6B & 8.00 & 4.19 & 14.59 & 19.63 & 15.23 & 10.97 & 31.22 & 55.15 & --$^*$ & --$^*$ & --$^*$ & --$^*$ & 20.65 & 14.26 & 39.79 & 61.74 \\ 
                
            \midrule
            
            \multirow{3}{*}{StarCoderBase} & 1B & 5.95 & 4.27 & 11.52 & 24.68 & 14.43 & 10.67 & 28.18 & 52.96 & 18.62 & 13.65 & 37.23 & 59.47 & \textbf{20.5} & \textbf{14.05} & \textbf{39.47} & \textbf{61.68} \\
            
             & 3B & 9.64 & 8.64 & 20.70 & 33.93 & 15.12 & 11.64 & 29.84 & 55.90 & 20.43 & \textbf{14.68} & 39.51 & 61.95 & \textbf{20.87} & 14.50 & \textbf{40.23} & \textbf{62.40} \\
             
             & 7B & 9.47 & 8.05 & 20.10 & 30.24 & 16.65 & 12.91 & 33.67 & 58.31 & --$^*$ & --$^*$ & --$^*$ & --$^*$ & 21.86 & 15.08 & 41.25 & 63.21 \\  
             
             \midrule
             
             ChatGPT & -- & 9.57 & 14.89 & 20.75 & 54.11 & 14.68 & 15.97 & 34.61 & 59.89 & --$^\dagger$ & --$^\dagger$ & --$^\dagger$ & --$^\dagger$ & --$^\dagger$ & --$^\dagger$ & --$^\dagger$ & --$^\dagger$ \\
             \bottomrule
        \end{tabular}
        \begin{tablenotes}
            \footnotesize
            \item $^*$ StarCoderBase-7B and CodeGen-Multi-6B are too large for effective fine-tuning.
            \item $^\dagger$ ChatGPT is a non-open source model and cannot be fine-tuned. We also cannot apply \ours{} to it.
        \end{tablenotes}
    \end{threeparttable}
\end{table*}

\noindent2)\textit{\;Results:} Table~\ref{tab:effectiveness_of_PromptCS} shows the performance of our \ours{} and baselines in terms of the four evaluation metrics, i.e., BLEU, METEOR, ROUGE-L, and SentenceBERT. Observe that task-oriented fine-tuning performs best on the LLM PolyCoder in terms of scores of the four metrics, followed by \ours{}, instruction prompting (few-shot), and instruction prompting (zero-shot). \ours{} outperforms zero-shot learning and few-shot learning, indicating that continuous prompts produced by the well-trained prompt agent are better than human-written prompts. The superior performance of task-oriented fine-tuning can be attributed to its approach of fully tuning model parameters on downstream task datasets, enabling the model to adapt to task-specific patterns, vocabulary, and objectives. This enhances relevance and accuracy while minimizing interference from irrelevant data. On the LLMs CodeGen-Multi and StarCoderBase, \ours{} is overall comparable to task-oriented fine-tuning, and both are significantly better than the two instruction prompting schemes. \ours{} even outperforms task-oriented fine-tuning on some LLMs, such as CodeGen-Multi-2B  (row 7), StarCoderBase-1B (row 9) and -3B (row 10). 
We also conduct instruction prompting schemes on a commercial LLM ChatGPT, and its performance is shown in the last row of Table~\ref{tab:effectiveness_of_PromptCS}. 
Observe that the performance of both instruction prompting schemes on ChatGPT is better than that on the other three LLMs in all four metrics, but is also not satisfactory compared to task-oriented fine-tuning and \ours{}. Although this comparison is unreasonable, we have reason to believe that if ChatGPT is open source, the task-oriented fine-tuning scheme and \ours{} will significantly improve its code summarization capabilities compared to instruction prompting schemes. 

\begin{table}[t]
    \centering
    \scriptsize
    \caption{LLMs' zero-shot learning performance using PI1 and PI2.}
    \label{tab:swap_prompt}

    \begin{threeparttable}
        \begin{tabular}{cccccccccc}
        \toprule
        \multirow{2}{*}{Base Model} & \multirow{2}{*}{Model Size} &   \multicolumn{2}{c}{BLEU} & \multicolumn{2}{c}{METEOR} & \multicolumn{2}{c}{ROUGE-L} & \multicolumn{2}{c}{SentenceBERT}\\

        \cmidrule{3-10}
        
        & & PI1 & PI2 & PI1 & PI2 & PI1 & PI2 & PI1 & PI2 \\
        
        \midrule

        \multirow{3}{*}{PolyCoder} & 160M & \textbf{7.98} & 4.72 & 3.40 & \textbf{3.76} & \textbf{13.82} & 5.99 & 16.12 & \textbf{38.17} \\
            
             & 0.4B & \textbf{7.18} & 6.80 & 2.67 & \textbf{5.54} & 11.55 & \textbf{12.33} & 16.89 & \textbf{38.03}\\
             
             & 2.7B & \textbf{7.68} & 4.99 & 2.28 & \textbf{3.95} & \textbf{12.23} & 7.40 & 12.92 & \textbf{35.06} \\ 
             
            \midrule
             
            \multirow{3}{*}{CodeGen-Multi} & 350M & \textbf{6.13} & 4.17 & \textbf{7.46} & 3.51 &\textbf{13.76} & 6.0 & \textbf{40.26} & 34.13\\
            
            & 2B & \textbf{7.96} & 3.14 & \textbf{3.08} & 2.55 & \textbf{13.42} & 5.16 & 15.26  & \textbf{23.94}\\
            
            & 6B & \textbf{8.00} & 3.89 & \textbf{4.19} & 3.50 & \textbf{14.59} & 7.48 & 19.63 & \textbf{26.46}\\ 
                
            \midrule
            
            \multirow{3}{*}{StarCoderBase} & 1B & 5.95 & \textbf{8.29} & 4.27 & \textbf{7.8} & 11.52 & \textbf{18.42} & 24.68 &  \textbf{36.00}\\
            
             & 3B & \textbf{9.64} & 6.95 & \textbf{8.64} & 6.42 & \textbf{20.70} & 15.4 & \textbf{33.93} & 27.75  \\
             
             & 7B & \textbf{9.47} & 5.57 & \textbf{8.05} & 4.70 & \textbf{20.10} & 12.36 & 3\textbf{0.24} & 20.86 \\ 
        
        \midrule

        ChatGPT & - & 3.39 & \textbf{9.57} & 12.48 & \textbf{14.89}  & 12.76 & \textbf{20.75} & 49.48 & \textbf{54.11}\\
        
        \bottomrule
        \end{tabular}
    \end{threeparttable}
    
\end{table}

To eliminate the threat of different prompts to the fairness of experimental results, we use both PI1 and PI2 for zero-shot learning on each LLM. The results are shown in Table~\ref{tab:swap_prompt}. It is observed that ChatGPT performs better when prompting with PI2. While for other LLMs, there is no unified conclusion about whether PI1 or PI2 is a more effective prompt, depending on the base LLM and the metric. This again reflects the difficulty of designing a high-quality prompt instructions. But in general, for LLMs except for ChatGPT, whether using PI1 or PI2, the performance of zero-shot learning does not exceed task-oriented fine-tuning and \ours{}. Thus, it does not threaten the rationality of our conclusion.

\summary{In adapting LLMs to code summarization tasks, \ours{} significantly outperforms instruction prompting with zero-shot and few-shot learning, and performs comparably to task-oriented fine-tuning. \ours{} even outperforms task-oriented fine-tuning on some LLMs, e.g., CodeGen-Multi-2B, StarCoderBase-1B and -3B.}

\begin{table*}[!t]
    \footnotesize
    \centering
    \tabcolsep=10pt
    \caption{Comparison of task-oriented fine-tuning and \ours{} in training and inference time.}
    \label{tab:compare_training_time_PromptCS_fine-tuning}
    \begin{tabular}{cccccc}
    \toprule
    \multirow{2}{*}{LLM} & \multirow{2}{*}{Model Size} & \multicolumn{2}{c}{Task-oriented Fine-tuning} & \multicolumn{2}{c}{\ours{}} \\

    \cmidrule(lr){3-4}\cmidrule(lr){5-6}
    
    & & Training Time & Inference Time & Training Time & Inference Time \\
    
    \midrule
  
    \multirow{3}{*}{CodeGen-Multi} & 350M & 20h:06m:44s & 02h:16m:03s & 15h:01m:39s & 00h:48m:45s \\
    
    & 2B & 81h:08m:24s & 09h:33m:44s & 58h:43m:29s & 04h:15m:24s \\
    
    & 6B & 151h:55m:57s$^*$ & 17h:26m:00s & 88h:03m:40s & 08h:34m:52s \\ 
    
    \midrule

    \multirow{3}{*}{PolyCoder} & 160M & 12h:25m:19s & 01h:09m:04s & 09h:41m:45s & 00h:26m:44s \\
    
    & 0.4b & 22h:18m:34s & 02h:25m:45s & 14h:32m:49s & 00h:50m:47s \\
  
    & 2.7b & 81h:54m:24s & 09h:24m:08s & 27h:41m:15s & 02h:12m:37s \\

    \midrule
    
    \multirow{3}{*}{StarCoderBase} & 1B & 50h:06m:52s & 03h:42m:50s & 29h:17m:32s & 01h:33m:52s \\
    
    & 3B & 82h:26m:21s & 08h:53m:41s & 43h:51m:44s & 02h:11m:46s \\
  
    & 7B & 211h:05m:18s$^*$ & 20h:58m:07s & 67h:21m:31s & 03h:56m:10s \\ 
    
    \bottomrule
    \end{tabular}
    \begin{tablenotes}
        \item $^*$StarCoderBase-7B and CodeGen-Multi-6B are too large for effective fine-tuning. Here is the single-epoch training time for it.
    \end{tablenotes}
\end{table*}

In addition, we also compare the training costs between the task-oriented scheme and \ours{} in terms of training time. Table~\ref{tab:compare_training_time_PromptCS_fine-tuning} lists the time costs incurred during training and inference on three LLMs. Observe that compared to the task-oriented scheme, \ours{} noticeably requires less time on all LLMs, which can be attributed to the fact that \ours{} is non-invasive to LLMs and does not update the parameters of the LLMs during training. It should be noted that when the model size of the LLM is very large (e.g., StarCoderBase-7B), the task-oriented fine-tuning will be very time-consuming and become unacceptable if the available training resources (e.g., the performance and number of GPU devices) are few. 
In terms of inference time, the inference time of \ours{} is much shorter than that of the corresponding task-oriented fine-tuned LLM. The reason behind this is that \ours{} is non-invasive to the parameters of LLMs, so LLMs' weights can be loaded with half precision (fp16 instead of fp32), which speeds up the inference and reduces GPU usage at the cost of only subtle accuracy loss. Hence, except for the inference time, the memory usage of \ours{} is also much less than the memory usage of task-oriented fine-tuning. For example, when the base LLM is PolyCoder-0.4b, fine-tuning the LLM requires about 11.5GB while \ours{} only takes up 2.5GB.

\summary{In terms of the training cost (e.g., training time, inference time, memory usage), \ours{} is notably lower than task-oriented fine-tuning, especially when the LLM has a larger model size.}

\begin{table*}[!t]
    \centering
    \footnotesize
    \tabcolsep=2.5pt
    \caption{Effects of prompt length and concatenation mode on \ours{}. $\mathcal{B}$: BLEU; $\mathcal{M}$: METEOR; $\mathcal{R}$: ROUGE-L; $\mathcal{S}$: SentenceBERT.}
    \label{tab:effects_of_prompt_length_concatenation_mode}
    
    \begin{tabular}{cccccccccccccc}
    \toprule
    
    \multirow{2}{*}{LLM} & \multirow{2}{*}{Prompt Length} & \multicolumn{4}{c}{Front-end Mode} & \multicolumn{4}{c}{Back-end Mode} & \multicolumn{4}{c}{Two-end Mode} \\
    
    \cmidrule(lr){3-6} \cmidrule(lr){7-10} \cmidrule(lr){11-14}
    
    & & $\mathcal{B}$ & $\mathcal{M}$ & $\mathcal{R}$ & $\mathcal{S}$ & $\mathcal{B}$ & $\mathcal{M}$ & $\mathcal{R}$ & $\mathcal{S}$ & $\mathcal{B}$ & $\mathcal{M}$ & $\mathcal{R}$ & $\mathcal{S}$ \\ 
    
    \midrule
    
    \multirow{5}{*}{StarCoderBase-1B} & 10 & 20.26 & 14.11 & 39.45 &  0.6169 & 20.19 & 13.99 & 39.36 &  0.6137 & 20.03 & 13.91 & 39.23 &  0.6163  \\
    
    & 20 & 20.40 & 14.03 & 39.47 &  0.6164 & 20.13 & \textbf{14.23} & 39.42 & 0.6165 & 20.29 & 14.17 & 39.50 & \textbf{0.6180} \\
    
    & 50 & \textbf{20.41} & 14.12 & 39.55 &  0.6185 & 20.36 & 14.07 & 39.48 &  0.6147 & \textbf{20.48} & 13.99 & \textbf{39.57} &  0.6153 \\
    
    & 100 & 20.30 & 14.16 & 39.53 & 0.6176 & \textbf{20.50} & 14.05 & 39.47 & \textbf{0.6168} & 20.07 & \textbf{14.26} & 39.37 &  0.6169  \\
    
    & 200 & 20.32 & \textbf{14.40} & \textbf{39.68} & \textbf{0.6206} & 20.27 & 14.16 & \textbf{39.56} &  0.6161 & 20.21 & 14.15 & 39.46 &  0.6165 \\
    
    \bottomrule
    \end{tabular}%
\end{table*}

\subsubsection{\textbf{RQ2. Influence of key configurations on \ours{}}}
\label{subsubsec:answer_to_RQ2}
\

As shown in Figure~\ref{fig:framework_of_our_approach}, two key configurations in the prompt agent may affect the performance of \ours{}, i.e., the prompt length that is determined by the pseudo prompt and the concatenation mode of the code embedding and prompt embedding. To reveal their effects, we conduct comprehensive experiments involving five different prompt lengths (including 10, 20, 50, 100, and 200) and three concatenation modes (including front-end mode, back-end mode, and two-end mode, detailed in Section~\ref{subsec:fusion_embedding_generation}). 
We uniformly utilize StarCoderBase-1B as the base LLM. The experimental results are present in Table~\ref{tab:effects_of_prompt_length_concatenation_mode}. 
From Table~\ref{tab:effects_of_prompt_length_concatenation_mode}, it is observed that if using the front-end mode, \ours{} achieves the best BLEU score at a prompt length of 50 (i.e., 20.41), while obtaining the best METEOR, ROUGE-L, and SentenceBERT scores at a prompt length of 200 (i.e., 14.40, 39.68, and 0.6206, respectively). If using the back-end mode, \ours{} achieves the best BLEU and SentenceBERT scores at a prompt length of 100, obtaining the best METEOR and ROUGE-L scores at prompt lengths of 20 and 200, respectively. If using the two-end mode, \ours{} obtains the best BLEU and ROUGE-L scores at a prompt length of 50, getting the best METEOR and SentenceBERT scores at prompt lengths of 100 and 20, respectively. These observations indicate that different combinations of two key configurations indeed have varying effects on the effectiveness of \ours{}. 
It is worth noting that although different combinations have varying effects on \ours{}, numerically, the differences in these effects are minimal. For example, in terms of BLEU, the best combination is the back-end mode and a prompt length of 100 (obtaining a score of 20.50), and the worst combination is the two-end mode and a prompt length of 10 (obtaining a score of 20.03). The difference in BLEU scores between the two combinations is less than 0.5. 

However, it should be noted that as the prompt length increases, the number of pseudo tokens needed to be trained also increases, which will consequently elevate the training cost of the model. Figure~\ref{fig:time_cost_with_different_prompt_length} shows how long it takes to train \ours{} for one epoch under different prompt length settings. From this figure, it can be intuitively observed that as the prompt length increases, the time cost required for training increases significantly.

\summary{Different combinations of the two key configurations have different effects on the effectiveness of \ours{}, yet the differences between these combinations are insignificant. However, it is important to note that increasing the prompt length will raise the training cost of the model.}

\begin{table}[!t]
    \caption{Performance of \ours{} when building the prompt encoder on different network architectures. 
    We bold the metric scores where the prompt encoder built on BiLSTM/Transformer performs better.
    }
    \label{tab:influence_of_prompt_encoder_on_different_models} 
    \centering
    \footnotesize
    \tabcolsep=3pt
    \begin{tabular}{cccccccccc}
        \toprule
        \multirow{2}{*}{LLM} & \multirow{2}{*}{Model Size} & \multicolumn{4}{c}{BiLSTM} & \multicolumn{4}{c}{Transformer} \\ 
        
        \cmidrule(lr){3-6} \cmidrule(lr){7-10}
        

         & & BLEU & METEOR & ROUGE-L & SentenceBERT & BLEU & METEOR & ROUGE-L & SentenceBERT \\ 
        
        \midrule

        PolyCoder & 160M & 16.01 & 11.68 & 34.04 & 0.5644 & \textbf{16.39} & \textbf{12.03} & \textbf{34.64} & \textbf{0.5688} \\ 
        
        CodeGen-Multi & 350M & \textbf{18.28} & 12.96 & \textbf{37.18} & 0.5920 & 18.09 & \textbf{12.99} & 37.09 & \textbf{0.5954} \\
        
        StarCoderBase & 1B & \textbf{20.50} & \textbf{14.05} & \textbf{39.47} & \textbf{0.6168} & 19.82 & 13.93 & 39.09 & 0.6111 \\ 
        
        \bottomrule
    \end{tabular}
\end{table}

\subsubsection{\textbf{RQ3. Influence of the network architecture used in the prompt encoder on \ours{}}}
\label{subsubsec:answer_to_RQ3}
\

As mentioned in Section~\ref{subsec:prompt_embedding_generation}, the prompt agent contains a DL-based prompt encoder built on BiLSTM. 
In practice, we also experiment to build the prompt encoder on a more advanced network, i.e., Transformer, to verify the impact of different choices on \ours{}. 
As in Section~\ref{subsubsec:answer_to_RQ1}, in this experiment, we uniformly set the prompt length to 100, and the concatenation mode to the back-end mode. Table~\ref{tab:influence_of_prompt_encoder_on_different_models} presents the experimental results. 

Observe that compared with building the prompt encoder on BiLSTM, 1) on PolyCoder-160M,  building the prompt encoder on Transformer brings performance improvements to \ours{} in all four metrics; 2) on CodeGen-Multi-350M, building the prompt encoder on Transformer enhances performance improvements to \ours{} in METEOR and SentenceBERT, but results in a decrease in BLEU and ROUGE-L; 3) on StarCoderBase-1B, building the prompt encoder on Transformer not only fails to improve \ours{}'s performance, but also causes performance degradation in all four metrics. 
Overall, as the model size of the LLM increases, the performance of \ours{} tends to improve if the prompt encoder is built on BiLSTM, while \ours{}'s performance shows the opposite trend if the prompt encoder is built on Transformer. 

\summary{Although Transformer appears to be more advanced compared to BiLSTM, experimental results in our application scenario do not show a significant advantage for Transformer. In other words, BiLSTM is sufficient to meet our needs for the prompt encoder design.}

\subsubsection{\textbf{RQ4. Influence of the training data size on \ours{}}}
\label{subsubsec:answer_to_RQ4}
\
\newline
In this paper, we also analyze the impact of training data size on the effectiveness of \ours{} and the task-oriented fine-tuning. To disclose this impact, we commence with a small-scale training set and systematically augment the number of training samples. The smaller training sets are randomly sampled from the complete training data. The considered training set sizes encompass 100, 1000, 10000, and 164923 (complete training data). In this experiment, we also uniformly utilize StarCoderBase-1B as the base LLM and set the prompt length and the concatenation mode to 100 and the back-end mode, respectively. 

\begin{figure}[!t]
    \centering
    \begin{minipage}[t]{0.3\linewidth}
        \includegraphics[width=\linewidth]{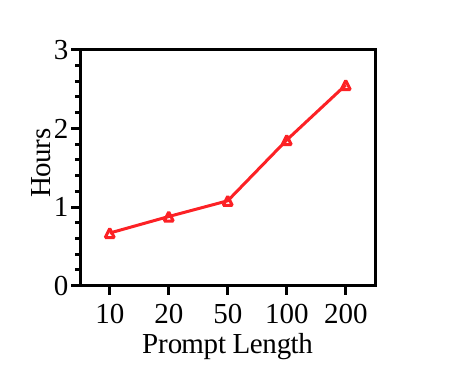}
        \caption{Training costs for different prompt lengths}
        \label{fig:time_cost_with_different_prompt_length}
    \end{minipage}
    \hfill
    \begin{minipage}[t]{0.3\linewidth}
        \centering
        \includegraphics[width = \linewidth]{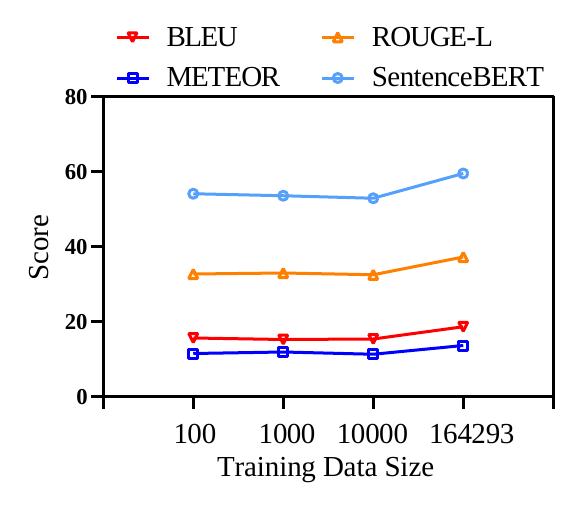}
        \caption{Influence of training set size on task-oriented fine-tuning}
        \label{fig:influence_of_training_set_size_finetune}
    \end{minipage}
    \hfill
    \begin{minipage}[t]{0.3\linewidth}
        \centering
        \includegraphics[width = \linewidth]{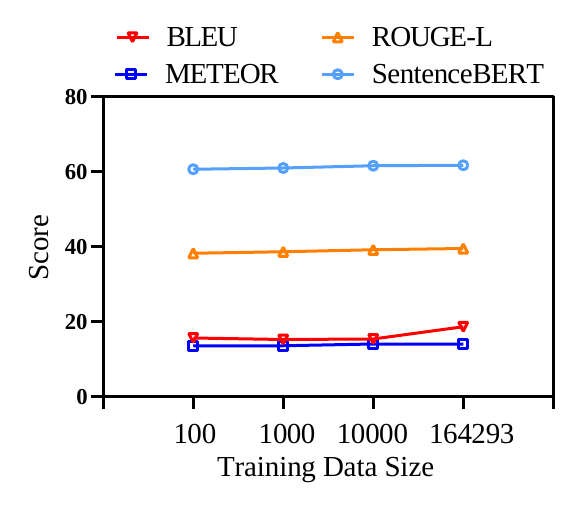}
        \caption{Influence of training set size on \ours{}}
        \label{fig:influence_of_training_set_size_promptcs}
    \end{minipage}
\end{figure}

Figure~\ref{fig:influence_of_training_set_size_finetune} and Figure~\ref{fig:influence_of_training_set_size_promptcs} showcase the experimental results, where the x-axis denotes the training set size, and the y-axes denote BLEU, METEOR, ROUGE-L, and SentenceBERT. Observe that the LLM performs significantly better when adopting the task-oriented fine-tuning on the complete dataset. Contrastly, with an increase in the size of the training set, the scores of \ours{} on the four metrics will increase, but the increase is not obvious. And \ours{} trained with 100 samples performs comparably to \ours{} trained with 164,923 samples. This underscores \ours{}'s superior adaptability and generalization capabilities on a small-scale dataset. It holds profound practical significance for reducing training costs, improving training efficiency, and obtaining satisfactory performance in environments with limited data.

\summary{While fine-tuning LLMs requires large datasets to achieve appreciable results, \ours{} can achieve decent performance even with limited training data resources, e.g., only 100 available samples.}

\begin{table*}[t]
    \scriptsize
    \caption{Effectiveness of \ours{} in different programming languages. $\mathcal{B}$: BLEU; $\mathcal{M}$: METEOR; $\mathcal{R}$:ROUGE-L; $\mathcal{S}$: SentenceBERT. 
    We bold the metric scores where Task-oriented Fine-tuning/\ours{} performs better.}
    \tabcolsep=1.4pt
    \label{tab:effectiveness_in_different_languages} 
    \begin{tabular}{cccccccccccccccccc}
        \toprule
        \multirow{4}{*}{LLM} & \multirow{4}{*}{Model Size} & \multicolumn{8}{c}{CSN-JavaScript} & \multicolumn{8}{c}{CSN-Python} \\
        
        \cmidrule(lr){3-10} \cmidrule(lr){11-18} 
        
        & & \multicolumn{4}{c}{Task-oriented Fine-tuning} & \multicolumn{4}{c}{\ours{}} & \multicolumn{4}{c}{Task-oriented Fine-tuning} & \multicolumn{4}{c}{\ours{}} \\ 

        \cmidrule(lr){3-6} \cmidrule(lr){7-10}\cmidrule(lr){11-14}\cmidrule(lr){15-18} 
        
        & & $\mathcal{B}$ & $\mathcal{M}$ & $\mathcal{R}$ & $\mathcal{S}$ & $\mathcal{B}$ & $\mathcal{M}$ & $\mathcal{R}$ & $\mathcal{S}$ & $\mathcal{B}$ & $\mathcal{M}$ & $\mathcal{R}$ & $\mathcal{S}$ & $\mathcal{B}$ & $\mathcal{M}$ & $\mathcal{R}$ & $\mathcal{S}$ \\

         
        \midrule

        PolyCoder & 160M & \textbf{14.28} & \textbf{9.03} & \textbf{25.87} & \textbf{51.33} & 13.49 &  8.26 & 25.39 & 48.67 & 11.78 & 6.36 & 22.34 & 28.96 & \textbf{11.85} & \textbf{7.17} & \textbf{23.38} & \textbf{30.81} \\ 
        
        CodeGen-Multi & 350M & 14.12 &  \textbf{10.30} & \textbf{28.11} & \textbf{52.70} & \textbf{14.48} & 9.98 & 28.07 & 51.79 & 12.20 & 7.56 & 24.20 & 34.91 & \textbf{12.36} & \textbf{8.45} & \textbf{25.37} & \textbf{36.20} \\

        StarCoderBase & 1B & \textbf{15.65} &  \textbf{10.71} & \textbf{30.29} & 53.92 & 15.62 & 10.60 & 30.03 & \textbf{54.45} & 12.96 & 8.73 & 26.17 & 37.36 & \textbf{13.88} & \textbf{9.20} & \textbf{27.85} & \textbf{41.03} \\

        \midrule

        \multicolumn{2}{c}{Average} & \textbf{14.68} & \textbf{10.01} & \textbf{28.09} & \textbf{52.56} & 14.53 & 9.61 & 27.83 & 51.64 & 12.31 & 7.55 & 24.24 & 33.75 & \textbf{12.70} & \textbf{8.27} & \textbf{25.53} & \textbf{36.01} \\
        
        \bottomrule
    \end{tabular}%
\end{table*}

\subsubsection{\textbf{RQ5. Effectiveness in other programming languages}}
\label{subsubsec:answer_to_RQ5}
\

To validate \ours{}'s generalization ability, we also conduct experiments in two other programming languages, including JavaScript and Python. In these experiments, we also uniformly set the prompt length and the concatenation mode to 100 and the back-end mode, respectively. The experimental results are shown in Table~\ref{tab:effectiveness_in_different_languages}.

From the CSN-JavaScript column of Table~\ref{tab:effectiveness_in_different_languages}, it is observed that the task-oriented fine-tuning scheme outperforms \ours{} on most metrics and LLMs. Nonetheless, it is worth noting that on three LLMs, \ours{} can achieve on average 98\% of the performance of the task-oriented fine-tuning scheme, which is encouraging. 
From the CSN-Python column of Table~\ref{tab:effectiveness_in_different_languages}, it is observed that on all three LLMs, \ours{} consistently performs better than the task-oriented fine-tuning scheme in all four metrics.

\summary{\ours{} exhibits good generalization ability on code summarization tasks in other different programming languages, including JavaScript and Python.}

\subsubsection{\textbf{RQ6. \ours{}'s performance in human evaluation}}
\label{subsubsec:human_evaluation}
\

In addition to automated evaluation, we conduct a human evaluation by following the previous works~\cite{2021-SiT, 2020-Hybrid-DeepCom, 2020-R2Com, 2020-Rencos, 2022-Automated-Human-Evaluation-Code-Documentation-Generation} to evaluate the summaries generated by three baselines and \ours{}. Specifically, we invite five volunteers with more than four years of software development experience and excellent English ability to carry out the evaluation. We randomly select 100 code snippets from the CSN-Java dataset, the corresponding ground-truth summaries, and summaries generated by baselines and \ours{}. Each volunteer is asked to assign scores from 1 to 5 to the generated summaries based on how similar they are to the corresponding ground-truth summaries, where 1 means ``Not Similar At All'' and 5 means ``Highly Similar/Identical''. To ensure the fairness and objectivity of experimental results, the order of the summaries in each case is shuffled and each summary is evaluated by five volunteers, and the final score is the median value of their scores.

\begin{table}[t]
    \centering
    \footnotesize
    \caption{Results of human evaluation. 
    }
    \label{tab:human_evaluation}
    
    \begin{tabular}{lccccccccc}
    \toprule
    Score & 1 & 2 & 3 & 4 & 5 & Avg. & $\geq 4$ & $\geq 3$ & $\geq 2$ \\
    
    \midrule
    
    Instruction Prompting with zero-shot learning & 41 & 11 & 27 & 18 & 3 & 2.31 & 21 & 48 & 52 \\
    
    Instruction Prompting with few-shot learning & 8 & 10 & 33 & 41 & 8 & 3.31 & 49 & 82 & 18 \\
    
    Task-oriented Fine-tuning. & 0 & 5 & 42 & 41 & 12 & 3.60 & 53 & 95 & 5 \\
    
    \ours{} & 0 & 7 & 37 & 41 & 15 & 3.64 & 56 & 93 & 7 \\
    \bottomrule
 \end{tabular}
 \vspace{-2mm}
\end{table}

Table~\ref{tab:human_evaluation} shows the score distribution of the generated summaries. Observe that \ours{} achieves the best scores and improves the average (Avg.) score from 2.31 (instruction prompting with zero-shot learning), 3.31 (instruction prompting with few-shot learning), and 3.60 (task-oriented fine-tuning) to 3.64. Specifically, among the randomly selected 100 code snippets, \ours{} can generate 15 highly similar or even identical summaries with the ground-truth ones (score = 5), 56 good summaries (score $\geq$ 4) and 93 summaries that are not bad (score $\geq$ 3). \ours{} also receives the smaller number of negative results (score $\leq$ 2). 
Based on the 100 final scores for each baseline and \ours{}, we follow~\cite{2020-Rencos} and conduct Wilcoxon signed-rank tests~\cite{1963-Wilcoxon} and compute Cliff’s delta effect sizes~\cite{2011-Cliffs-Delta-Calculator}. Comparing \ours{} with instruction prompting with zero-shot learning, instruction prompting with few-shot learning, and task-oriented fine-tuning, the p-values of Wilcoxon signed-rank tests at 95\% confidence level are 5.46E-16, 0.016, and 0.722, which means the improvements achieved by \ours{} are statistically significant over instruction prompting with zero-shot and few-shot learning. In addition, Cliff's delta effect sizes are 0.5669 (large), 0.1478 (small), and 0.0314 (negligible), respectively.

\summary{Human evaluation shows that the summaries generated by \ours{} can achieve higher scores on average compared to those generated by baselines. \ours{} also generates the largest number of good summaries.}
\section{Case Study}
\label{sec:case_study}
In this section, we provide two code summarization cases to understand the generated summaries of \ours{} compared with baselines. Both cases are real-world examples from the CSN-Java test set. For each case, we consider the comment of the code snippet as the ground-truth summary and generate summaries by applying baselines and \ours{} to the LLM StarCoderBase-3B.

\begin{figure}[t]
  \centering
  \includegraphics[width=0.9\linewidth]{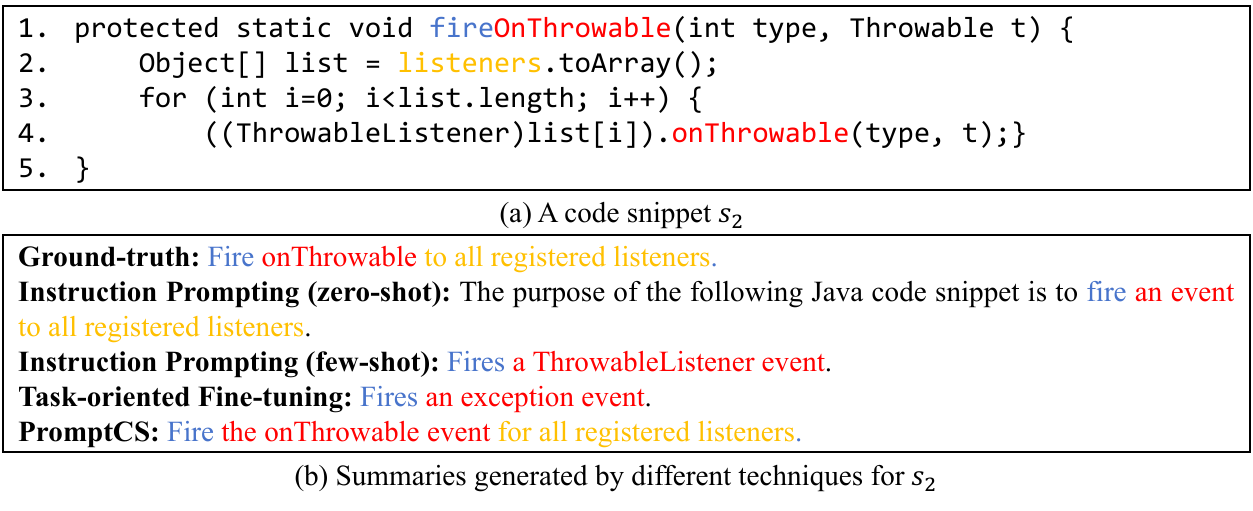}
  \caption{Code summarization case 1}
  \label{fig:discussion_case_1}
\end{figure}

Figure~\ref{fig:discussion_case_1} shows the first code summarization case, where (b) presents the ground-truth summary and the summaries generated by baselines and \ours{} for the code snippet $s_2$ in (a). From the figure, it is observed that compared with the ground-truth summary, the summaries generated by instruction prompting (few-shot) and task-oriented task fine-tuning can cover the semantics of the first two parts, i.e., ``Fire'' (Blue font) and ``onThrowable'' (Red font), whereas the summaries generated by instruction prompting (zero-shot) and \ours{} can cover the semantics of all three parts (including ``to all registered listeners'' (Orange font)). However, upon closer inspection of the method name of $s_2$ shown in the first line of Figure~\ref{fig:discussion_case_1}(a), we can find that for the second part (i.e., ``onThrowable''), the summaries generated by three baselines (i.e., ``an event'', ``a ThrowableListener event'', and ``an exception event'', respectively) are inaccurate. Only \ours{} successfully induces the LLM to generate this part correctly.

\begin{figure}[htbp]
  \centering
  \includegraphics[width=0.9\linewidth]{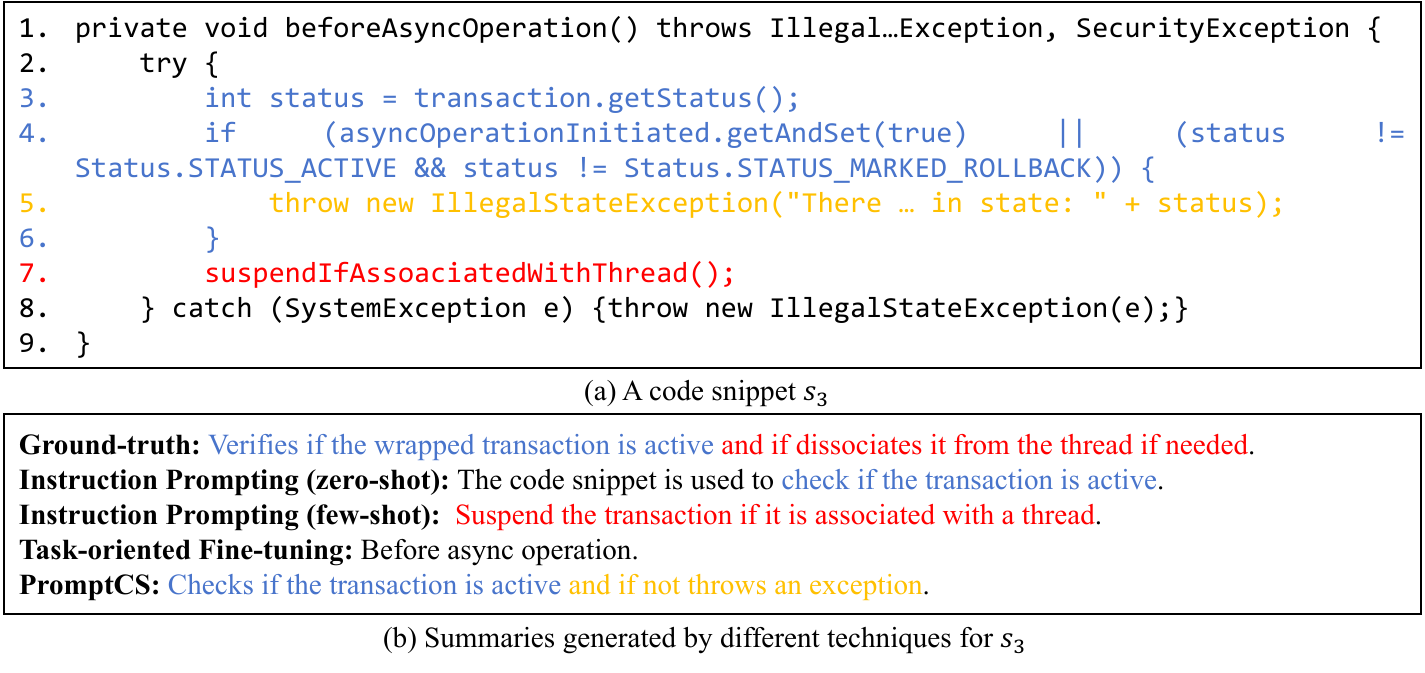}
  \caption{Code summarization case 2}
  \label{fig:discussion_case_2}
\end{figure}

Figure~\ref{fig:discussion_case_2} shows the second code summarization case, where the ground-truth summary shown in (b) can be split into two clauses: ``Verifies if the wrapped transaction is active'' and ``if dissociates it from the thread if needed''. Observe that compared with the ground-truth summary, 1) the summaries generated by instruction prompting (zero-shot and \ours{} cover the first clause; 2) the summary generated by instruction prompting (few-shot) covers the second clause; 3) the task-oriented fine-tuning fails to cover any of clauses. Another noteworthy point is that although the summary ``if not throws an exception'' generated by \ours{} is not part of the ground-truth summary, it indeed provides an excellent summary for the statement \texttt{throw new IllegalStateException("There ... in state: " + status);} of the code snippet $s_3$.

From the two cases above, it is apparent that compared to baselines, \ours{} performs better in adapting LLMs for code summarization tasks. The credit for \ours{}'s good performance goes to its prompt agent, which is trained under the guidance of LLMs, allowing it to generate prompts better suited for LLMs. Of course, the second case also shows that \ours{} still has room for improvement. We will further optimize it in the future to better induce LLMs to generate high-quality code summaries.

\section{Related Work}
\label{sec:related_work}

\subsection{Code Summarization}
\label{subsec:code_summarization_related_work}
Research on code summarization started more than a decade ago. Most early code summarization techniques~\cite{2010-Program-Comprehension-with-Code-Summarization, 2010-Towards-Generating-Summary-Java-Methods, 2010-Automated-Text-Summarization-Summarizing-Code, 2013-Automatic-Generation-Summaries-for-Java-Classes} are extractive methods. Such methods work by extracting a subset of the statements and keywords from the code, and then including information from those statements and keywords in the final generated summary. 
For instance, Sonia Haiduc et al.~\cite{2010-Program-Comprehension-with-Code-Summarization} first propose an extractive method to generate summaries for code snippets. 
Subsequently, in light of the widespread availability of extensive code datasets and the success of neural machine translation (NMT) research in the field of NLP~\cite{2014-GRU, 2015-NMT-Jointly-Learning-to-Align-Translate}, deep learning approaches have increasingly become prevalent in enhancing the capabilities of code summarization tools. Pioneers in the field have discovered that certain Seq2Seq models, including RNN~\cite{2014-RNN-Encoder-Decoder} and LSTM~\cite{2016-CODE-NN}, demonstrate the ability to effectively capture the semantic relationships between code snippets and their corresponding summaries. For example, Iyer et al.~\cite{2016-CODE-NN} employ LSTM networks with attention to generating sentences describing C\# code snippets and SQL queries. Hu et al.~\cite{2018-DeepCom} combine SBT with LSTM to automatically generate code comments for Java methods. However, it is noteworthy that RNN-based methodologies encounter challenges in maintaining long-term dependencies. To overcome this limitation and address potential bottlenecks in modeling long sequences, some Transformer-based methods have been proposed. Zhang et al.~\cite{2020-Rencos} propose a retrieval-based neural source code summarization approach by enhancing a Transformer-based model with the most similar code snippets retrieved from the training set. Ahmad et al.~\cite{2021-SiT} explore the effectiveness of using Transformer to capture long-range dependencies in code for generating code summaries. 
Recently, there are also a lot of studies that apply LLMs to code summarization tasks~\cite{2024-LLM4CodeSum}. For example, Ahmed et al.~\cite{2022-Few-shot-Training-LLMs-for-Code-Summarization} investigate the effectiveness of few-shot training in adapting LLMs to code summarization and find that it can make Codex significantly outperform fine-tuned pre-trained language models, such as CodeT5. Given the concern of potential code asset leakage when using commercial LLMs (e.g., GPT-3.5), Su et al.~\cite{2024-Distilled-GPT-for-Code-Summarization} utilize knowledge distillation technique to distill small models from LLMs. Their experimental findings reveal that the distilled small models can achieve comparable code summarization performance to LLMs. Gao et al.~\cite{2023-What-Makes-Good-In-Context-Demonstrations} investigate the optimal settings for in-context learning, including few-shot example selection methods, few-shot example order, and the number of few-shot examples. They find that carefully designed few-shot examples can significantly improve LLMs' performance on code-related tasks (including code summarization tasks). Geng et al.~\cite{2024-LLM-Few-Shot-Summarizers-Multi-Intent-Comment-Generation} reveal the ability of LLMs to address multi-intent comment generation. Ahmed et al.~\cite{2024-Semantic-Augmentation-of-Prompts-for-Code-Summarization} propose to enhance few-shot samples with semantic facts automatically extracted from the source code. Sun et al.~\cite{2023-Automatic-Code-Summarization-via-ChatGPT} design some heuristic questions to collect the feedback of ChatGPT, thereby finding an appropriate prompt to guide ChatGPT to generate in-distribution code summaries. In our work, we train a prompt agent under the guidance of LLMs to generate continuous prompts. Our method is non-invasive to LLMs and freezes the parameters of LLMs when training the prompt agent, which can greatly reduce the requirements for training resources. 

\subsection{LLM for SE}
\label{subsec:llm_for_se}
With the success of LLMs in NLP, a large number of SE researchers have started integrating them into the resolution process of various SE tasks~\cite{2023-Survey-on-LLMs-for-SE, 2023-LLM-for-SE}. For example, Han et al.~\cite{han2024archcode} leverages LLMs combined with in-context learning to interpret software requirements from textual descriptions for code generation. Jain et al.~\cite{jain2022jigsaw} explore the integration of LLMs (e.g., Codex) in generating code from natural language descriptions of programmer intent. Dakhel et al.~\cite{dakhel2023effective} utilize mutation testing to augment the prompts, guiding LLMs to generate test cases that can detect software bugs. Sun et al.~\cite{sun2024gptscan} attempt to detect smart contract logic vulnerabilities by combining GPT and static analysis. 
In this paper, we focus on adapting LLMs for code summarization tasks~\cite{2022-Few-shot-Training-LLMs-for-Code-Summarization, 2023-Automatic-Code-Summarization-via-ChatGPT, 2023-Chatgpt-Programming-Assistant}. 

\noindent\textbf{Adapting LLMs to SE with PEFT.} As mentioned in Section~\ref{sec:introduction}, SE researchers have conducted many empirical studies to investigate the effectiveness of PEFT methods from NLP in adapting LLMs to code-related tasks. For instance, 
Wang et al.~\cite{2022-No-More-Fine-tuning-in-Code-Intelligence} conduct an empirical study to evaluate the usage and effect of prompt tuning in code intelligence tasks. They conduct prompt tuning on CodeBERT and CodeT5 and experiment with three code intelligence tasks including defect prediction, code summarization, and code translation. The intuition of prompt tuning is to convert the training objective of downstream tasks into a similar form as the pre-training stage, i.e., the MLM objective~\cite{2019-BERT, 2020-CodeBERT}. They find that prompt tuning outperforms fine-tuning and is more effective in low-resource scenarios.  
Martin et al.~\cite{2023-Parameter-Efficient-Fine-Tuning-Code-Generation} investigate the effectiveness of PEFT techniques in automated code generation. They find that LLMs with PEFT consistently and significantly outperform small language models under the same GPU limit. 
Liu et al.~\cite{2024-PEFT-in-Code-Change-Learning} examine the effect of two PEFT methods, namely Adapter Tuning (AT)~\cite{2021-Adapter-Tuning-for-Pretrained-LM-Adaptation} and Low-Rank Adaptation (LoRA)~\cite{2022-LoRA} on code-change-related tasks, including just-in-time defect prediction~\cite{2021-Just-In-Time-Defect-Prediction} and commit message generation~\cite{2014-Generating-Commit-Messages-via-Summarization}.
Liu et al.~\cite{2023-Empirical-Study-of-PEFT-for-Code-Models} conduct an empirical study of PEFT methods (including Adapter~\cite{2019-Parameter-Efficient-Transfer-Learning-for-NLP}, LoRA~\cite{2022-LoRA}, Prefix tuning~\cite{2021-Prefix-Tuning}, and MHM~\cite{2022-Unified-View-of-PEFT}) in various code understanding and code generation tasks. 
Zou et al.~\cite{2023-PEFT-on-SE-Tasks} examine the effectiveness of five PEFT methods, including Houlsby (i.e., Adapter~\cite{2019-Parameter-Efficient-Transfer-Learning-for-NLP}), Pfeiffer~\cite{2020-MAD-X}, Parallel (i.e., MHM~\cite{2022-Unified-View-of-PEFT}), Prefix tuning, and LoRA, on eight models and four code-related tasks, including clone detection, defect detection, code search, and code translation. 
Zhuo et al.~\cite{2024-Astraios} investigate 5 tasks and 8 different datasets encompassing both code comprehension and code generation tasks on StarCoder~\cite{2023-StarCoder}. They find that PEFT generally leads to the best downstream performance, and PEFT methods differ significantly in their efficacy based on the model scale. 
Although these empirical studies cover various PEFT methods and a range of code-related tasks, they merely report the effectiveness of applying PEFT methods to adapt LLMs for code-related tasks. They lack in-depth analysis of the performance of specific PEFT methods and the factors influencing their effectiveness on particular code-related tasks. Unlike them, our \ours{} is an effective and customized prompt learning framework for the code summarization task. In addition, we delve deeply into the factors that may impact \ours{}' performance, including prompt encoder architecture design, prompt length, and concatenation modes. 

\section{Conclusion}
\label{sec:conclusion}
We propose a prompt learning framework \ours{} for source code summarization. \ours{} is equipped with a prompt agent that can induce LLMs to accomplish code summarization tasks by generating continuous prompts. \ours{} can save model users from spending a significant amount of time crafting prompt instructions. Comprehensive automated and human evaluations demonstrate that \ours{} is effective in adapting LLMs to code summarization with low training costs. We believe that our prompt learning framework paired with LLMs can also benefit other SE tasks, and we leave the exploration of other SE tasks to future work.

\Acknowledgements{
The authors at Nanjing University were supported, in part by the National Natural Science Foundation of China (61932012 and 62141215). The authors at Nanyang Technological University are supported by the National Research Foundation, Singapore, and DSO National Laboratories under the AI Singapore Programme (AISG Award No: AISG2-GC-2023-008).
}

\bibliographystyle{unsrt}
\bibliography{reference}

\end{document}